\documentclass{aa}
\usepackage{psfig}
\begin{document}

   \thesaurus{(08.19.5;   % SNe individual (SN 1998bw)
               13.07.1;  % Gamma rays: bursts,
               13.18.2)  % Radio continuum: general,
              }
\def\SN{SN\thinspace{1998bw}}
\def\GRB{GRB\thinspace{980425}}

\authorrunning{Wieringa et al.}
\titlerunning{SN\thinspace{1998bw}}

   \title{SN\thinspace{1998bw}: The Case for a Relativistic Shock}

   \author{M. H. Wieringa\inst{1}, S. R. Kulkarni\inst{2}, \& 
     D. A. Frail\inst{3}
          }
 
   \offprints{D. A. Frail}
 
   \institute{{Australia Telescope National Facility, CSIRO,
              Epping 2121, Australia,
                                     }
   \and {Division of Physics, Mathematics, and Astronomy 105-24,
               California Institute of Technology, Pasadena, CA 91125 USA}
   \and {National Radio Astronomy Observatory (NRAO), 
               Socorro, NM 87801 USA
               email: dfrail@nrao.edu}}
 
   \date{Received ; accepted }

   \maketitle

\begin{abstract}
\SN\ shot to fame by claims of association with \GRB.
Independent of its presumed association with a GRB,  this SN is unusual
in its radio properties. A simple interpretation of the unusually
bright radio emission leads us to the conclusion that there are two
shocks in this SN: a slow moving shock containing most of the ejecta
and a relativistic shock ($\Gamma=2$) which is responsible for the
radio emission.  This is the first evidence for the existence of
relativistic shocks in supernovae.   It is quite plausible  that this
shock may produce high energy emission (at early times and by inverse
Compton scattering).  As with other supernovae, we expect radio
emission at much later times powered primarily by the slow moving
ejecta. This expectation has motivated us to continue monitoring this
unusual SN.
\end{abstract}

%
%________________________________________________________________

\section{Introduction}

Accounts of the discovery of optical and radio emission from the
supernova SN\thinspace{1998bw} has been given elsewhere (Galama et al.
1998, Kulkarni et al. 1998). The primary link between \SN\ and \GRB\ is
the coincidence of the two events in space and time (Galama et al.
1998). Unfortunately, the coincidence window (16-arcmin error circle,
$\pm$ 1 day) is not sufficiently small to make a firm claim of the
association. 

BeppoSAX NFI observations show that there are two sources, a secularly
fading source coincident (within the NFI error radius of 1.5 arcmin)
with \SN\ and a source which is not coincident with the SN and which
appeared to fade by the end of the first NFI observation (initiated 10
h after the burst and total duration of 40 h). The latter source has
not been detected in subsequent NFI observations (see GCN 151 for
summary; also contribution by E. Pian in these proceedings).  A simple
interpretation is that the transient source is the x-ray afterglow of
\GRB. Another possibility is that the transient source is unrelated to
\GRB. Upcoming ASCA observations may clarify this confusing situation.

What has been lost in all the exciting developments is that \SN\ is
{\it an unusual supernovae, regardless of its association to \GRB}.
Indeed, it's claimed and controversial association tends to overshadow
the unusual nature of this SN. We now exclusively focus on the radio
properties of this SN.  The radio observations were conducted with the
Australia Telescope Compact Array (ATCA)\footnote{The Australia
  Telescope is funded by the Commonwealth of Australia for operation
  as a National Facility managed by CSIRO.} as a part of our GRB effort.

\section{Radio Observations}

In Figure 1 we present the up-to-date radio light curve of \SN.  The
radio luminosity of \SN\ is unusually high but what really sets it
apart from other radio SN is the emergence of copious radio emission
at early times.  Supernovae expand and thus their size increases with
time. Thus the parameter which best distinguishes \SN\ from other
radio SN is the specific intensity (i.e. the ratio of flux to the
solid angle of the source). It is conventional in radio astronomy to
express specific intensity in units of brightness temperature and the
conversion is done using the Rayleigh-Jeans formula.

We assume that the radio emission originates from the same region as
the optical emission in which case the expansion speed is 60,000 km
s$^{-1}$ The predicted angular expansion is then $\sim 1 \mu$arcsec
per day. Such a source would be compact enough, particularly in the
first two weeks, to suffer from strong scattering due to density
inhomogeities in the Galactic interstellar medium. In contrast to
GRB\thinspace{970508} (Frail et al. 1997), only a smooth rise to a
maximum was seen for the radio emission from SN\thinspace{1998bw}.
This strongly suggests that the expansion of the radio photopshere
greatly exceeds that of the optical. Kulkarni et al.  (1998) inferred
$v_{exp}>0.3c$ from the absence of refractive scintillation at 20 and
13 cm.

As noted in Kulkarni et al. (1998) the peak brightness temperature of
\SN\ is $10^{13}$ K. This is two orders of magnitude higher than that
inferred for previously studied radio SN (Chevalier 1998). The
inferred brightness temperature is also in excess of the well known
inverse Compton limit $T_{\rm icc}\sim{5}\times{10}^{11}$ K for a
source radiating via the incoherent synchrotron mechanism (Kellermann
\& Pauliny-Toth 1968, Readhead 1994).

As pointed out by Readhead (1994), high brightness temperatures also
result in exceedingly large estimates of minimum energy.  The total
energy of a synchrotron source is $$ U/U_{eq}\ = \ 1/2\ \eta^{11}(1\ +
\ \eta^{-17}),$$ where $\eta=\theta/\theta_{eq}$ and $\theta_{eq}$ is
referred to as the equipartition angular radius and $U_{eq}$ is the
equipartition energy which is also (approximately) the minimum energy.
The strong dependence of $U$ on $\eta$ consequently means that a high
price must be paid in $U$ for sources smaller than, or larger than
$\theta_{eq}$. As noted by Kulkarni et al. (1998), if the angular size
used is consistent with $v_{exp}$=60,000 km s$^{-1}$ then
$\theta=7\times\theta_{eq}$, and therefore the source energy would be
dominated by relativistic electrons U$_e={10}^{54}$ erg - much larger
than the total energy release in a typical supernova. Therefore, the
only reasonable hypothesis is to assume $\theta\simeq\theta_{eq}$,
leading to v$_{exp}$=1.2c-1.9c, U$_{eq}\simeq{5}\times{10}^{48}$ erg,
and M$_{ej}\simeq{10}^{-5}$ M\sun.

%                                                One column figure
%----------------------------------------------------------- S_vib
%\begin{figure}
%\vspace{1.01cm}
%\hspace{0cm}\psfig{figure=smlog_tb.ps,width=8.8cm}
%\vspace{-.51cm}
%      \caption[]{The evolution of the brightness temperature of
%        SN\thinspace{1998bw}. Four wavelengths, 20 cm (cross), 13 cm
%        (star), 6 cm (circle) and 3 cm (square) are plotted together.
%        Time (x-axis) is determined assuming that the SNe and the
%        gamma-ray burst are related. The y-axis is the brightness
%        temperature for a distance of 38 Mpc and a velocity of
%        $v=60,000v_{60}$ km s$^{-1}$.}
%         \label{tbee}
%   \end{figure}
%

\section{An Alternative Model}

Recently, Waxman \&\ Loeb (1998) have produced an alternative {\it
  sub-relativistic} model to explain the radio curve shown in Figure
1.  In their model, the radio emission arises in the post-shocked gas
(shock speed of 60,000 km s$^{-1}$). They assume rapid equilibration
between the electrons and protons and a modest compressed magnetic
field.  In this model, the energy in the magnetic field is many orders
of magnitude smaller than that in the energetic particles.  The
curious result is that Waxman \&\ Loeb (1999) are able to reproduce
the observed spectrum (save for a high frequency point) and an energy
estimate comparable to that of Kulkarni et al. This is curious because
it is a general result that the total energy is minimized only when
there is equipartition between the electrons and magnetic field
strength.

The Waxman \&\ Loeb  analysis is based on the assumption of
mono-energetic electrons. Sari, Kulkarni \&\ Phinney (1999)
have carried out a detailed analysis using a thermal energy spectrum and
find that for a non-equipartition plasma (as envisaged by Waxman \&\
Loeb) the total energy is much higher than $10^{49}$ erg. Thus it
appears that the simplification used by Waxman \&\ Loeb does grossly
underestimate the total energy. Thus we conclude that either the energy
of the  radio emitting plasma is $10^{52}$ erg or that there exists
a relativistic shock in this SN.

\section{The Future: Evidence For A Slow Shock?}

We have argued above that the radio emission in the first $\sim$100
days had its origin in a mildly relativistic shock which carries only
a small amount of mass (10$^{-5}$ M\sun) and energy (10$^{49}$ erg).
The bulk of the ejecta mass and energy of the SN is presumably traced
by the optical photosphere. Thus, it is reasonable to expect some
late-time radio emission from this slower-moving shock as it interacts
with any circumstellar material. We continue to make multi-wavelength
measurements toward SN\thinspace{1998bw} at the ATCA with this in
mind. Our results to date shown in Figure 1 suggests that the
power-law-like decay has persisted for at least 250 days.

%                                                One column figure
%----------------------------------------------------------- S_vib
   \begin{figure}
\vspace{1.01cm}
\hspace{0cm}\psfig{figure=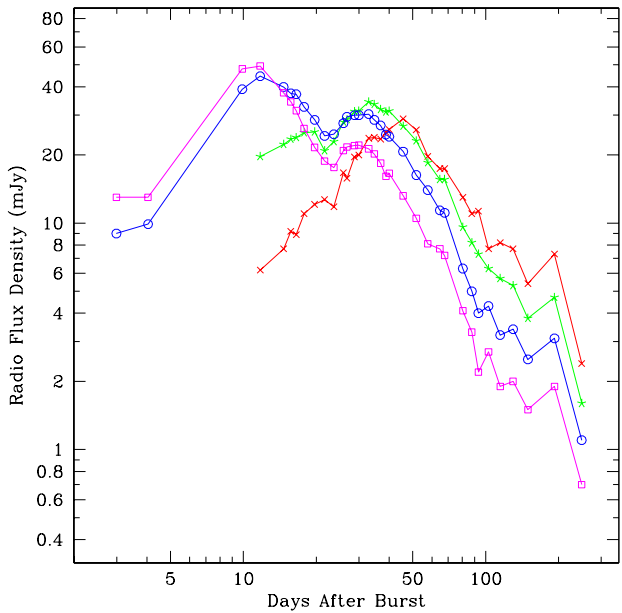,width=8.8cm}
\vspace{-.51cm}
      \caption[]{The radio light curve of SN\thinspace{1998bw} at four
        wavelengths (20cm=cross, 13cm=star, 6cm=circle and
        3cm=square). The last measurements were made on 1998 December
        30, nearly 250 days after the gamma-ray burst. On this day the
        1-$\sigma$ uncertainty in these flux measurements is 0.5 mJy
        for 20 and 13-cm, and 0.3 mJy for 6 and 3-cm.}
         \label{G980703}
   \end{figure}
%
%______________________________________________________________

%\begin{acknowledgements}
%  The Australia Telescope is funded by the Commonwealth of Australia
%  for operation as a National Facility managed by CSIRO.  The NRAO is
%  a facility of the National Science Foundation operated under
%  cooperative agreement by Associated Universities, Inc. DAF thanks
%  Bodan Paczy\'nski for useful discussions and his continued
%  encouragement of the ATCA monitoring effort.
%\end{acknowledgements}

\end{document}